\documentclass[12pt]{article}

\usepackage[english]{babel}
\usepackage[utf8]{inputenc}
\usepackage{titlesec}
	\titleformat*{\section}{\normalfont}
	\titleformat*{\subsection}{\normalfont}
	\titleformat*{\subsubsection}{\normalfont}

\usepackage{fancyhdr}
\usepackage[a4paper,top=2.54cm,bottom=2.54cm,left=2.54cm,right=2.54cm,marginparwidth=1.75cm]{geometry}

\usepackage{amsmath}
\usepackage{graphicx}
\usepackage[colorinlistoftodos]{todonotes}
\usepackage[colorlinks=true, allcolors=blue]{hyperref}
\usepackage{indentfirst}

\begin{document}
\begin{titlepage}
	\thispagestyle{fancy}
	{\centering

	{\huge\bfseries SeQUeNCe GUI: An Extensible User Interface for Discrete-Event Quantum Network Simulations\par}
		\vspace{1.5cm}
	{\Large Alexander Kiefer\par}
    \vspace{2cm}
    {\large Office of Science, Science Undergraduate Laboratory Internship Program \par
	Indiana University, Bloomington, IN \par}
	\vspace{1cm}
	{\large Argonne National Laboratory
    \par Lemont, Illinois
	\par}
	\vfill
	{\large August, 2021\par}}
	\vfill
    
	Prepared in partial fulfillment of the requirement of the Department of Energy, Office of Science’s Science Undergraduate Laboratory Internship Program under the direction of Martin Suchara at the Argonne National Laboratory. \par
	\vspace{2cm}
    Participant: Alexander Kiefer
    \par
    \vspace{2cm}
    Research Advisor: Martin Suchara
    \vfill
\end{titlepage}

\begin{abstract}
\setcounter{page}{2}
With recent advances in the fields of quantum information theory [J. Pablo. Nature 12, 2172 (2021)]\nocite{ataides2021xzzx} and the approach of the Noisy Intermediate-Scale Quantum (NISQ) [J. Preskill. Quantum 2, 79 (2018)]\nocite{preskill2018quantum} computing era, it is necessary to provide tools for experimentation and prototyping that are able to keep pace with the rapidly progressing field of quantum computing. SeQUeNCe, an open source simulator of quantum network communication, aims to provide scalability and extensibility for the simulation of quantum networks, from the hardware level to the application and protocol level. In order to improve upon the usability of this software, we implement a graphical user interface which maintains the core principles of SeQUeNCe, scalability and extensibility, while enhancing the software’s portability and ease of use. We demonstrate the capabilities of the graphical user interface through the construction of the existing Chicago metropolitan quantum network topology.
\end{abstract}

\section{INTRODUCTION}

Quantum computing promises advances in specific applications over classical computers. This theoretical improvement, known as quantum advantage, is the primary motivator of quantum information research today. In more simple terms, pioneering physicist Richard Feynman was quoted as saying, “Nature isn't classical, dammit, and if you want to make a simulation of nature, you'd better make it quantum mechanical, and by golly it's a wonderful problem, because it doesn't look so easy.” 

Since Feynman’s time, this observation has been one of the primary drivers of research in the field of quantum information theory. Though a demonstrated quantum advantage has been proven theoretically in a number of algorithms, we have yet to reach the stage of rigorously demonstrating these advantages experimentally due to the current state of quantum computing. More specifically, we are currently limited by the low number of qubits and relatively high noise to signal ratios in present hardware. 

In order to surpass the physical limitations of quantum computers, researchers rely on quantum simulation. This allows quantum algorithms and protocols to be run on classical computers. Though this presents the disadvantage of simulating actual physical quantum properties like entanglement, superposition, and teleportation, rather than using them natively, it allows researchers to test and prototype their work without the difficulties that are present on current physical quantum hardware.

Quantum networking is one specific sub domain of quantum computing which has benefited tremendously from the use of simulators. With current quantum computers being highly experimental, expensive, and largely isolated, it is exceedingly difficult to conduct experiments on physical quantum networks. SeQUeNCe\cite{wu2020sequence}, or Simulator of QUantum Network Communication, is a customizable, discrete-event quantum network simulator which aims to provide researchers an extensible and open source means of developing components for quantum networks and the future quantum internet.

In order to improve upon the simulator, we develop a graphical user interface for SeQUeNCe which maintains the core principles of customizability and open software architecture, while increasing simulation portability and ease of use.

\section{DESIGN PRINCIPLES}

The development of quantum networks and the quantum internet are still in their nascent stages, with a proposed framework by the Internet Engineering Task Force (IETF) \cite{van-meter-qirg-quantum-connection-setup-01} currently being the most well defined. In order to accommodate a variety of different network stack architectures, SeQUeNCe is abstracted into six primary modules, those being Application, Network Management, Entanglement Management, Resource Management, Hardware, and the Simulation Kernel. This allows users to specify entirely new components on the network stack, from hardware to application layer. In order to maintain this level of modularity, the GUI component of SeQUeNCe must allow for this same level of modularity, as well as providing users with greater ease of use. We aim to accomplish this through the use of templates and JSON serialization.

\begin{figure}
    \centering
    \includegraphics[width=0.5\textwidth]{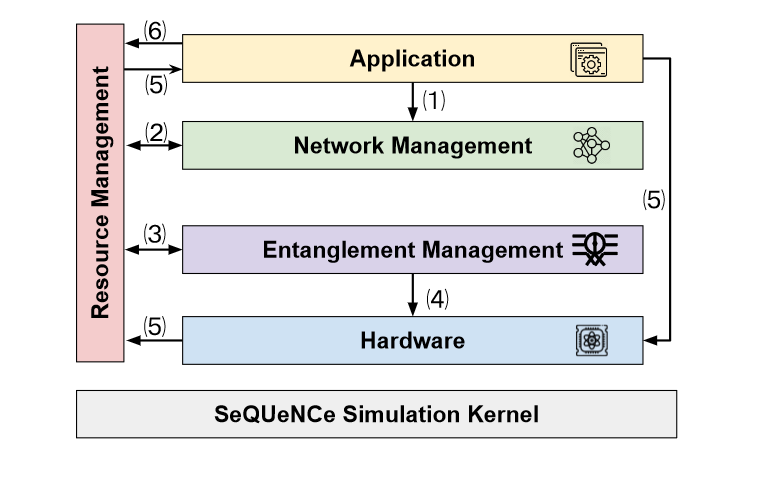}
    \caption{Architecture diagram of SeQUeNCe modules}
    \label{fig:arch}
\end{figure}

\section{DEVELOPMENT TOOLS}

The SeQUeNCe simulator is implemented natively within python. This allows for simple dependency management and compatibility with a wide variety of systems. In order to maintain this simplicity, we develop the GUI elements of sequence using packages which are also native to python. With a number of well supported GUI packages such as tkinter \cite{grayson2000python} and PyQt5 \cite{willman2021overview}, the selection of a specific framework was somewhat challenging. In order to provide the best cross-platform support, we decided on a web based framework called Dash. This framework manages React elements within python and allows one to take full advantage of more widely used JavaScript visualization libraries. By using a system of HTML templates and function callbacks, web apps can directly interface with the back-end simulation processing in Python.

\section{NETWORK GRAPH}

The graphical user interface implements a network graph using the widely supported Cytoscape.js \cite{franz2016cytoscape} library. Cytoscape.js is a standalone JavaScript library with no required dependencies and optional support for additional extensions. The specific implementation used for the GUI is a React component extension made to work specifically with the Dash visualization framework. For additional features like flow, cycle detection, and clustering, we manage the data of the network internally with NetworkX \cite{hagberg2008exploring}, a robust and optimized graph library implemented within python.

The network is visualized as an undirected graph, with each edge representing a dual channel connection between nodes, where one channel transmits classical information, and the second quantum. This network structure is preferred as information which does not have to be stored in a quantum state can be transmitted more simply as a classical signal, reducing the amount of bandwidth utilization on quantum links. 
The layout of the network is based upon the compound spring embedder (CoSE) algorithm \cite{karaccelik2012improved}. This is a physics based model which generates layouts by calculating optimal spring forces between nodes. The advantage of this algorithm is that it is has a low computational complexity on the quadratic order of |V|, where v is the number of vertices in the graph, and, it allows for the use of compound nodes, a feature intended to be implemented in the future for more complex network support.

The graph also contains a legend which maps the color of nodes to their respective types for ease of use. The legend is automatically updated to display only node types which are present in the current version of the graph.

\begin{figure}
    \centering
    \includegraphics[width=0.5\textwidth]{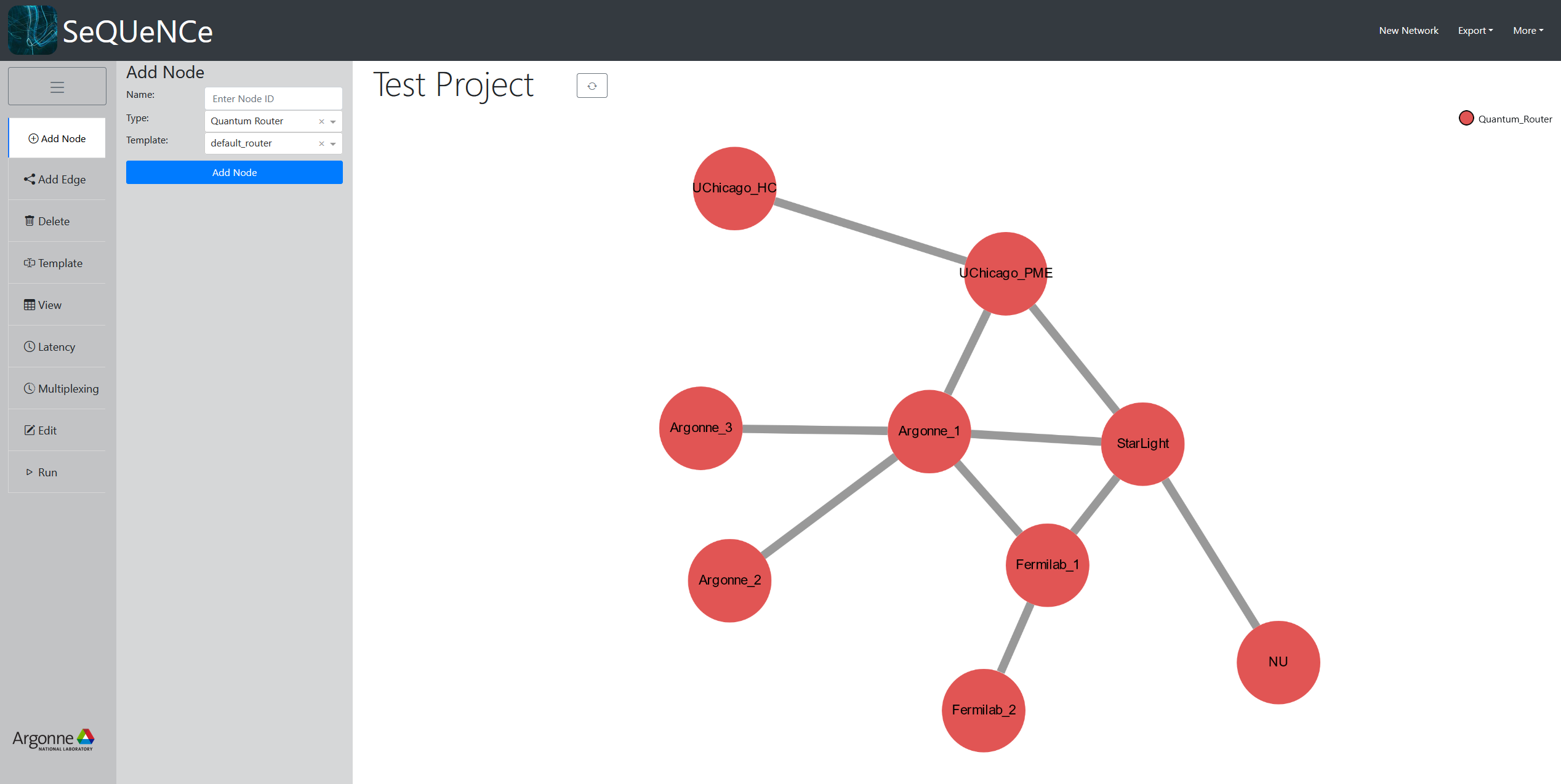}
    \caption{The Graphical User Interface (GUI) for SeQUeNCe}
    \label{fig:gui}
\end{figure}

\section{HARDWARE COMPONENTS}

\subsection{Quantum Memory}

Memories are modeled as single atom memory arrays, where each memory in the array is defined by its coherence time, frequency, efficiency, and fidelity. Coherence time describes the length of time that an entanglement operation can exist on a memory. The frequency describes the time needed for a memory to return to its ground state after excitation. Efficiency is the probability with which a photon is emitted from the memory. Finally, fidelity describes the quality of entanglement for a memory. Memory arrays are considered homogeneous and will not function properly otherwise.

\subsection{Quantum Router}

Routers are nodes which contain and manage memory arrays on the network. Each operates its own resource and network manager to keep track of operations and activity of its memories. In the current implementation of SeQUeNCe, any connection between two quantum routers implicitly creates a Bell State Measurement node between router nodes.

\subsection{Bell State Measurement}

Bell State Measurement (BSM) nodes are used to create maximally entangled particles and to allow for quantum teleportation. By measuring two qubits, this node determines which of the four Bell states the two qubits are in. If the qubits were not previously in Bell states, they are projected into one. This is the process by which qubits are entangled. This node is defined by the detector type contained within it.

\subsection{Photon Detector}

Photon detectors are measurement instruments used within BSM nodes to entangle qubits. Photon detectors are defined by the efficiency, count rate, dark count rate, and resolution. The efficiency is the probability that a photon is detected when it reaches the detector. The count rate is the inverse of the constant time taken after each measurement before a new measurement can be taken. The dark count rate is the number of false positive detections of photons due to outside interference. Finally, the resolution is the accuracy with which a detection can be made, in regard to time.

\section{USAGE}

The SeQUeNCe GUI can be separated into three primary categories, with additional miscellaneous features, such as table views of network elements. 

\subsection{Network Manipulation}

These features enable a user to interact with their network. By adding nodes, edges, deleting elements, or editing them, a user can specify the properties of components on the graph. When adding a node, a user can select the node name, type, as well as a template which corresponds to that type of node. All nodes in a network must have a unique name, with no duplicates allowed.

Edges do not use templates, as their properties are largely static and only consist of two values, those being distance and attenuation. Distance affects the propagation delay as

\begin{equation}
    L/c, 
\end{equation}

where \textit{L} is the distance and \textit{c} is the speed of light in the fiber. Attenuation affects the loss rate of the quantum channel as

\begin{equation}
    10-\frac{L \cdot \alpha_{0}}{10}, 
\end{equation}

where $\alpha_{0}$ is the attenuation measured in dB/km.

When elements are removed or added to the network, UI elements, such as the legend and table views, are appropriately updated. When using the edit menu, users can select elements of the graph to analyze and change using drop-down and input menus. These menus are also responsive and update themselves based on other selections. For example, if editing a node of the type “Quantum Router”, by changing the type to “Detector”, a user will only be able to select templates of the type “Detector”.

For the manipulation of latency and time-division multiplexing, the GUI utilizes data tables in the form of a graph adjacency matrix. By changing the values in this data table, one can adjust the latency on specific classical channels within the network, as well as configure multiplexing on quantum channels.

\subsection{Templating}

One of the key features of the SeQUeNCe GUI is the templating feature. This method of interaction allows users to specify any number of parameters for any type defined within the simulator. These templates can then be applied to the nodes of a network to define their parameters. Templates can be applied any number of times to any number of nodes and allow for simple reusability. Every template is defined using a unique id, along with the node type it defines. However, each template type specifies its own unique set of parameters.

\begin{figure}
    \centering
    \includegraphics[width=0.5\textwidth]{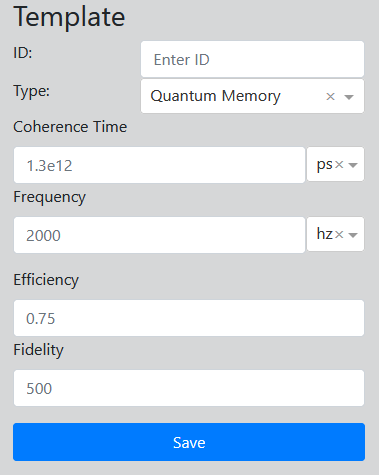}
    \caption{The templating menu}
    \label{fig:temp}
\end{figure}

\subsection{Simulation}

 Currently, the GUI provides simulation capability for random request application networks. This type of simulates what is considered regular network traffic across the currently configured network. The user can specify a name for the simulation as well as the duration of its runtime. The simulation will then run with an accompanying runtime and progress counter. Once complete, the simulation is saved to the SeQUeNCe install directory under the user given name. The default results output for such a simulation are the wait times, reservations, and throughput on every node within the network.

\begin{figure}
    \centering
    \includegraphics[width=0.5\textwidth]{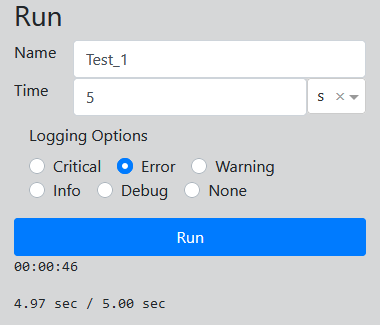}
    \caption{The simulation menu}
    \label{fig:sim}
\end{figure}

\section{JSON SERIALIZATION}

In order to allow for portability and reproducibility of simulations, the GUI is capable of exporting designed networks to JSON formatted files. There are three specific files which the GUI is capable of generating, those being topology, template, and simulation files. The topology JSON contains all of the information necessary to construct the physical layout of the network, including nodes, edges, and network properties such as latency and time-delay multiplexing. The template JSON contains the configuration settings for all defined templates. Templates in this file map directly to values contained within the topology file and are necessary to reconstruct a network. Finally, the simulation JSON contains information about the configuration of a specific simulation, allowing users to share simulations like benchmarks. This file is optional and currently still experimental in its implementation.

\section{ACKNOWLEDGEMENTS}

This work was supported in part by the U.S. Department of Energy, Office of Science, Office of Workforce Development for Teachers and Scientists (WDTS) under the Science Undergraduate Laboratory Internship (SULI) program.

\newpage
\bibliographystyle{ieeetr}
\bibliography{sample}
\end{document}